\documentclass[12pt]{article}

\usepackage{authblk}
\usepackage{amsthm,amsmath}
\RequirePackage[square,numbers]{natbib}
\usepackage{bm}
\usepackage{graphicx}
\usepackage[nopostdot,acronym,nomain,nonumberlist]{glossaries}
\usepackage{url}
\usepackage[margin=1in]{geometry}

\makeglossaries

\newacronym{aids}{AIDS}{Acquired Immunodeficiency Syndrome}
\newacronym{ari}{ARI}{Adjusted Rand Index}
\newacronym{art}{ART}{Antiretroviral Therapy}
\newacronym{crf}{CRF}{Circulating Recombinant Form}
\newacronym{urf}{URF}{Unique Recombinant Form}
\newacronym{dna}{DNA}{Deoxyribonucleic Acid}
\newacronym{haart}{HAART}{Highly Active Antiretroviral Therapy}
\newacronym{hiv}{HIV}{Human Immunodeficiency Virus}
\newacronym{mcmc}{MCMC}{Markov Chain Monte Carlo}
\newacronym{mh}{MH}{Metropolis-Hastings}
\newacronym[longplural={Men who have Sex with Men}]{msm}{MSM}{Men who have Sex with Men}
\newacronym{se}{SE}{Standard Error}
\newacronym{sd}{SD}{Standard Deviation}
\newacronym{sir}{SIR}{Susceptible-Infected-Recovered}
\newacronym{who}{WHO}{World Health Organization}
\newacronym{phi}{PHI}{Primary HIV Infection}
\newacronym{er}{ER}{Erdos-Renyi}
\newacronym{ws}{WS}{Watts-Strogatz}
\newacronym{ba}{BA}{Barabasi-Albert}
\newacronym{jc69}{JC69}{Jukes-Cantor 1969}
\newacronym{k80}{K80}{Kimura 1980}
\newacronym{tn93}{TN93}{Tamura-Nei 1983}
\newacronym{hky85}{HKY85}{Hasegawa-Kishino-Yano 1985}
\newacronym{gtr}{GTR}{General Time Reversible}
\newacronym{pr}{PR}{Protease}
\newacronym{rt}{RT}{Reverse Transcriptase}
\newacronym{nni}{NNI}{Nearest-Neighbour Interchange}
\newacronym{spr}{SPR}{Subtree Pruning and Regrafting}
\newacronym{cp}{CP}{ClusterPicker}
\newacronym{shcs}{SHCS}{Swiss HIV Cohort Study}
\newacronym[longplural={Sexually-Transmitted Infections}]{sti}{STI}{Sexually-Transmitted Infection}
\newacronym{pam}{PAM}{Percent Accepted Mutation}
\newacronym{wag}{WAG}{Whelan and Goldman}
\newacronym{jtt}{JTT}{Jones, Taylor, and Thornton}
\newacronym{nj}{NJ}{Neighbour-Joining}
\newacronym{wpgma}{WPGMA}{Weighted Pair-Group Method of Analysis}
\newacronym{smc}{SMC}{Sequential Monte Carlo}
\newacronym{is}{IS}{Importance Sampling}
\newacronym{sis}{SIS}{Sequential Importance Sampling}
\newacronym{em}{EM}{Expectation-Maximization}
\newacronym{aic}{AIC}{Akaike Information Criterion}
\newacronym{bic}{BIC}{Bayesian Information Criterion}
\newacronym{dp}{DP}{Dirichlet Process}
\newacronym{pyp}{PYP}{Pitman-Yor process}
\newacronym{tasp}{TasP}{Treatment as Prevention}
\newacronym{mrca}{MRCA}{Most Recent Common Ancestor}
\newacronym{map}{MAP}{Maximum Posterior probability}
\newacronym{ml}{ML}{Maximum Likelihood}
\newacronym{idu}{IDU}{Injection Drug User}

\glossarystyle{list}
\glsaddall

\begin{document}

\title{Transmission clusters in the HIV-1 epidemic among men who have sex with men in Montreal, Quebec, Canada}

\author[1]{Luc Villandr\'{e}\thanks{\textit{Corresponding author}: luc.villandre@mail.mcgill.ca}}
\affil[1]{Dept. of Epidemiology, Biostatistics, and Occ. Health\\
	  McGill University, Montreal, QC, Canada}

\author[2]{Aur\'{e}lie Labbe}
\affil[2]{Dept. of Decision Sciences\\
	  HEC Montr\'{e}al, Montreal, QC, Canada}
	  
\author[3]{Ruxandra-Ilinca Ibanescu}
\affil[3]{McGill AIDS Centre\\
	  Lady Davis Institute, Jewish General Hospital, Montreal, QC, Canada}
	  
\author[3]{Bluma Brenner}
	  
\author[4,5]{Michel Roger}
\affil[4]{Centre de Recherche du Centre Hospitalier de l'Universit\'{e} ́de Montr\'{e}al (CRCHUM)\\
  Montreal, QC, Canada}
\affil[5]{D\'{e}partement de microbiologie, infectiologie et immunologie\\
	  Universit\'{e} de Montr\'{e}al, Montreal, QC, Canada}
	  
\author[6]{David A. Stephens}
\affil[6]{Dept. of Mathematics and Statistics\\
	  McGill University, Montreal, QC, Canada}
\maketitle

\begin{abstract}

  \textbf{Background.} Several studies have used phylogenetics to investigate \gls{hiv} transmission among \glspl{msm} in Montreal, Quebec, Canada, revealing many transmission clusters. The Quebec \gls{hiv} genotyping program sequence database now includes viral sequences from close to $4,000$ \gls{hiv}-positive individuals classified as \glspl{msm}. In this paper, we investigate clustering in those data by comparing results from several methods: the conventional Bayesian and maximum likelihood-bootstrap methods, and two more recent algorithms, DM-PhyClus, a Bayesian algorithm that produces a measure of uncertainty for proposed partitions, and the Gap Procedure, a fast distance-based approach. We estimate cluster growth by focusing on recent cases in the \gls{phi} stage. \textbf{Results.} The analyses reveal considerable overlap between cluster estimates obtained from conventional methods. The Gap Procedure and DM-PhyClus rely on different cluster definitions and as a result, suggest moderately different partitions. All estimates lead to similar conclusions about cluster expansion: several large clusters have experienced sizeable growth, and a few new transmission clusters are likely emerging. \textbf{Conclusions.} The lack of a gold standard measure for clustering quality makes picking a best estimate among those proposed difficult. Work aiming to refine clustering criteria would be required to improve estimates. Nevertheless, the results unanimously stress the role that clusters play in promoting \gls{hiv} incidence among \glspl{msm}.
  

\end{abstract}

\section{Introduction}

  \glsreset{msm}
  \glsreset{phi}
  \glsreset{ari}
  \glsreset{haart}
  \glsreset{mcmc}
  \glsreset{map}

The genotyping of pathogens provide novel opportunities to improve understanding of epidemic dynamics, and as a result, phylogenetic models have become a common tool in the study of infectious disease transmission \cite{Dudas2014, Kenah2016, Foley2015, Huerta-Cepas2014}. Those models have been used extensively to study \gls{hiv} epidemics \cite{Brenner2013, Brenner2013a}, mainly due to the availability of large sequence databases, collected mainly in the context of antiretroviral drug resistance testing \cite{SHCS2010, LosAlamos, Hue2004}. The Quebec \gls{hiv} genotyping program database \cite{Brenner2007} for example, as of 2017, contains $27,487$ sequences from $9,687$ \gls{hiv}-positive individuals, living mostly in Montreal, Quebec, Canada. 

\glspl{msm} remain especially at risk of contracting \gls{hiv}: in Montreal, prevalence in that risk group could be as high as $13$\% \cite{Mtrack2013}. Phylogenetic analyses of sequences obtained from \glspl{msm} in the Quebec \gls{hiv} genotyping program database have revealed the existence of many large transmission clusters, and highlighted their association with incidence: $42$\% of \glspl{msm} infected between 2012 and 2015 belonged to a transmission cluster of size $20$ or more, compared to $13$\% \cite{Brenner2017} between 2004 and 2007. \gls{haart} has been successful in substantially suppressing viremia within the diagnosed population, making late transmission of the virus a lot less common, and consequently, early transmission has been increasingly driving the epidemic \cite{Charest2014}. Recently-infected individuals are much more likely to transmit because of high viral load, viral homogeneity, and immature immune response, potentially leading to \textit{transmission cascades}, consecutive transmission events happening in a short time span \cite{Brenner2013a}. Those cascades result specifically in the formation of transmission clusters, and increased clustering may therefore point to a higher proportion of early transmissions. 

Montreal has become a UNAIDS Fast Track City in May 2017, and the only way to reach the 90-90-90 targets \cite{UNAIDS2014} is to break the cycle of large cluster transmissions. Also, quantifying the role of early transmission in the epidemic is important from a public health standpoint, as it can help assess the extent to which programs are able to reach infected individuals early enough. This is the motivation behind the current study, in which we analyse a large sample of sequences collected via the Quebec \gls{hiv} genotyping program with a variety of clustering methods, comparing their estimates to shed light on the ongoing evolution of clustering in the epidemic. 

\subsection{Background}

Phylogenetic studies of clustering in \gls{hiv}-1 epidemics tend to rely on a number of \textit{ad hoc} rules applied a posteriori to phylogenetic estimates. Availability of software like MEGA and PAUP* \cite{Tamura2013, Swofford2003} has led to widespread adoption of maximum likelihood phylogenetic reconstruction, coupled with the bootstrap to evaluate confidence in the inferred clades. In that context, cluster estimation relies on an arbitrary cutoff applied to bootstrap support estimates, usually between $70$\% and $95$\% \cite{Hillis1993, Holder2003, Brenner2007}. Alternatively, software like BEAST and MrBayes \cite{Drummond2012, Ronquist2012} have popularised Bayesian phylogenetic estimation. Both are based on versions of the \gls{mcmc} algorithm, that numerically approximate posterior distributions for a variety of evolutionary and phylogenetic parameters. They also provide posterior probability support for clades, a crucial measure for the identification of clusters, which in phylogenetic terms correspond to non-nested clades forming a partition of the sample. For example, many studies require posterior probability support of $1$ to conclude in clustering \cite{Yang2006}.

In addition to clade confidence requirements, many studies also impose a within-cluster genetic distance requirement, usually between $0.01$ nt/bp and $0.05$ nt/bp \cite{Ragonnet-Cronin2013}. Distance requirements may be applicable to mean \cite{Hue2004}, median \cite{Prosperi2011}, or maximum \textit{patristic distances} \cite{Brenner2007}, that is, distances calculated by summing branch lengths along the shortest path between any two tips in the phylogeny. The \textit{ClusterPicker} algorithm \cite{Ragonnet-Cronin2013} instead formulates that requirement in terms of maximum within-cluster $p$-distances, e.g. the Hamming distance.

Cutoffs are however hard to justify rigorously \cite{Chalmet2010} and so, methods grounded in more explicit definitions of clusters have been published. \cite{Vrbik2015} proposed the so-called \textit{Gap Procedure}, a fast pure distance-based approach that requires minimal tuning. In a similar vein, \cite{Villandre2017arxiv} formulated DM-PhyClus, a Bayesian algorithm that aims to minimise reliance on thresholds while still offering rigorous inference for cluster membership.

The heavy computational burden of conventional phylogenetic inference is problematic in light of the fast increase in the size of sequence databases, and can therefore limit its applicability \cite{Wang2014}. Thankfully, software designed to handle larger datasets is now available. RAxML \cite{Stamatakis2014} and FastTree \cite{Price2010}, for example, make use of heuristics in phylogenetic optimisation to improve scalability of the maximum likelihood phylogenetic methods. Clustering of large datasets in a purely Bayesian paradigm is a computational challenge that has not yet been fully overcome, although vast progress has been made thanks in part to GPU computing \cite{Drummond2012, Ronquist2012}.

\subsection{Objectives}

This paper aims to provide up-to-date estimates of transmission clusters in the \gls{hiv} epidemic among \glspl{msm} in Montreal and assess their temporal expansion. In doing so, we seek to improve previous assessments of the contribution of early transmission and transmission cascades to the epidemic. We therefore perform an exhaustive clustering analysis of \gls{hiv}-1 subtype B sequences in the most recent version of the Quebec \gls{hiv} genotyping database originating from \gls{msm} subjects. There is a lack of consensus as to how clustering of \gls{hiv}-1 sequence data should be done, and different methods may produce equally valid, but conflicting results \cite{Brenner2013}. To assess sensitivity of cluster estimates to phylogenetic assumptions and cluster definitions, we compare results from a number of methods,
\begin{enumerate}
\item Maximum likelihood phylogenetic reconstruction, coupled with a bootstrap support requirement for clades, which we refer to as the \textit{ML-bootstrap} approach, \label{enum:MLclus}
\item Bayesian phylogenetic inference, coupled with a posterior probability support requirement for clades, \label{enum:BayesClus}
\item DM-PhyClus \cite{Villandre2017arxiv},
\item Gap Procedure \cite{Vrbik2015}.
\end{enumerate}

\section{Materials and methods}

\subsection{Data}

Among the entries in the Quebec \gls{hiv} genotyping program database \cite{Brenner2007}, we retain $3,936$ subtype B sequences, each obtained from a different individual classified as \gls{msm}. All sequences were collected between May 3, 1996 and February 1, 2016. We focus on a particular viral DNA genomic region, and study $918$ loci, covering sites $10$-$297$ of the protease region (PR), and $112$-$741$ of the reverse transcriptase (RT) region, of the \textit{pol} gene. Each sequence comes with a time stamp, indicating when the blood sample was collected, and an indicator of infection status, either chronic treated, chronic untreated, or \gls{phi}. A case is considered a \gls{phi} if the sequence was obtained less than $6$ months after seroconversion \cite{Brenner2007}. 

Since the analyses focus on transmission clusters among \glspl{msm} only, we exclude $20$ sequences obtained from women, leaving us with $3916$ sequences. To avoid potential artefacts resulting from selective pressure induced by antiretroviral therapy, we remove sequences for chronic treated patients and patients with missing infection status information as well, leaving us with $3,704$ sequences. Finally, for rooting purposes, we add to those $3,704$ sequences three subtype A outgroup sequences from Zambia \cite{LosAlamos} (NCBI accession numbers AB254141, AB254142, AB254143), and it follows that the data to analyse include $3,707$ sequences.

\subsection{Methods}

\subsection{Conventional maximum likelihood}

We obtain the maximum likelihood (ML) phylogenetic estimate \cite{Felsenstein1981} with RAxML 8.2.10 \cite{Stamatakis2014}, under the assumption that nucleotide evolution follows the GTR + I + $\Gamma(5)$ model. We produce $1,000$ bootstrap trees, and use them to evaluate confidence in clades present in the ``best scoring'' phylogeny, RAxML's estimate of the ML phylogeny. To conclude in clustering, we require, in turns, bootstrap support greater than $70$\%, $90$\%, or $95$\% and consider genetic distance requirements of $1.5$\%, $3$\%, $4.5$\%, $6.8$\%, and $7.7$\%. More specifically, we apply, in turns, the maximum within-cluster Hamming distance requirement of ClusterPicker \cite{Ragonnet-Cronin2013}, the maximum median within-cluster patristic distance requirement of PhyloPart \cite{Prosperi2011}, and the maximum within-cluster patristic distance requirement of \cite{Brenner2007}. For the PhyloPart analysis, \cite{Prosperi2011} recommend setting cutpoints based on percentiles of the total tree patristic distance distribution. In their real data analysis, they use the $15$th and $30$th percentiles as cutpoints, which we also try.

\subsection{Conventional Bayes}

We perform phylogenetic inference with MrBayes 3.2.6 \cite{Ronquist2012} using default parameters, under the assumption that nucleotide evolution follows the GTR + I + $\Gamma(4)$ model. MrBayes uses the \gls{mcmc} algorithm \cite{Hastings1970}, more specifically the so-called \textit{Metropolis-coupled \gls{mcmc}}, or (MC)$^3$ \cite{Geyer1991}, algorithm, to generate estimates for the posterior distribution of phylogenetic parameters. The \gls{mcmc} algorithm lets us recursively obtain samples from the posterior distributions of interest. It starts off by setting all parameters at an arbitrary value. Then, in each iteration, updates to parameter values are proposed, conditional on their current values. Each proposal is randomly accepted with probability equal to the Metropolis-Hastings (MH) ratio, producing a move in the parameter space; else, no move is recorded. After a large number of iterations, parameter values generated throughout the chain are used to empirically estimate the posteriors. We run three million iterations, burning in the first $50$\% and sampling one iteration out of $500$. We derive the majority rule consensus tree from the remaining $3,000$ trees, and produce cluster estimates by identifying clades with posterior probability support of $1.0$. Once again, we use the ClusterPicker algorithm to obtain cluster estimates, under the requirement that within-cluster distance be, in turns, bounded above by $1.5$\%, $3$\%, and $4.5$\%, $6.8$\%, and $7.7$\%.   

\subsection{Cutpoint selection} 

All conventional clustering approaches require selection of genetic distance and confidence cutpoints. Prior to the analyses, researchers involved directly in the Quebec HIV genotyping program performed a preliminary clustering of the dataset. The proposed partition results from successive updates, performed at different points in time, of an initial cluster estimate, obtained under the restriction that once a sequence is attributed to a cluster, it cannot be re-attributed to another cluster in a following analysis. They identified from the results seven noteworthy sets of sequences, comprising $372$ sequences in total, that they expect correspond with genuine transmission clusters. One of those sets, for instance, comprises $68$ sequences and is characterised by more than half of its members harbouring the Non-Nucleoside Reverse Transcriptase Inhibitors (NNRTI) mutation K103N. As in \cite{Prosperi2011}, we use that subsample as a reference set. We compare partitions obtained across a range of cutpoints with that set using the \gls{ari}, a measure of similarity between two partitions, with the aim of maximising overlap. Greater \gls{ari} values are better, and the measure is bounded above by $1$, indicating perfect correspondence. We describe the comparison scheme in more details in Supplementary Material S1.  

\subsection{DM-PhyClus}

DM-PhyClus is a Bayesian phylogenetic algorithm that aims to estimate transmission clusters directly, by identifying sets of sequences supported by distinctive subtrees, thus avoiding the need to specify thresholds arbitrarily \cite{Villandre2017arxiv}. Unlike conventional methods, as a way to directly find sets of sequences resulting from transmission cascades, it defines a cluster as a clade supported by a phylogeny with a distinctive branch length distribution, usually with a relatively small mean. Conditional on an input phylogeny -- the maximum likelihood estimate in this study -- it uses the \gls{mcmc} approach to produce an estimate of the posterior distribution of cluster membership indices. As a result, it has the added benefit of providing a straightforward measure of uncertainty for cluster and cluster membership estimates, in the form of co-clustering frequencies across the chain. It requires specification of a number of other priors and evolutionary parameters, which we list in Supplementary Material S3. We perform $220,000$ iterations, discarding the first $20,000$ as a burn-in and applying a thinning ratio of $1$ over $200$, leaving us with a sample of size $1,000$. We identify the partition that maximises the joint posterior probability score, which we refer to as the \gls{map} estimate.

\subsection{Gap Procedure}

The Gap Procedure is a pure distance-based clustering algorithm that requires minimal tuning, and avoids reliance on ad hoc cutpoints by partitioning sets of sequences into distinctive components without requiring phylogenetic estimation \cite{Vrbik2015}. When the true clusters are compact and separable enough, the Gap Procedure can propose partitions that largely agree with conventional phylogenetic estimates, but in a fraction of the time normally required for such analyses, thus making the method ideal for handling large datasets. For example, in an analysis presented in \cite{Vrbik2015}, partitioning a dataset comprising $627$ sequences of length $810$ took $126$ hours with MrBayes and less than a second with the Gap Procedure. The method takes as input a matrix of pairwise distances, which we obtain under the \gls{k80} model. We leave tuning parameters at their default values.

\subsection{Cluster growth evaluation}

To evaluate cluster growth properly, we would need to know seroconversion dates for all cases whose sequences were sampled. The dataset however contains instead an infection stage indicator, equivalent to a censored estimate of infection time, i.e. smaller (greater or equal) than six months prior to the sampling date for \glspl{phi} (chronic cases). Most \gls{hiv}-positive individuals are diagnosed while already in the chronic stage, at which point seroconversion date estimates are very imprecise \cite{Kouyos2011}. As a result, we use \glspl{phi} only to obtain a lower bound estimate for the growth of inferred clusters, since \glspl{phi} can be reliably associated with a short time window prior to sampling. We focus on a period ranging from January 1, 2012 to February 1, 2016. For example, one of the methods may propose a cluster of size $20$, with eight of its sequences having been obtained from cases diagnosed while in the \gls{phi} stage at some point in 2014. We can therefore be certain that those cases were infected after January 1, 2012 and so, we conclude that the cluster has accrued at least eight new cases in the selected period.



\subsection{Software}

Under the conventional methods, we get cluster estimates by importing phylogenetic estimates from RAxML or MrBayes into R v3.2.3 and analysing them with functions in the \textit{phangorn} and \textit{ape} libraries \cite{Schliep2011}. We use functions in the \textit{GapProcedure} and \textit{DMphyClus} R libraries to obtain the other estimates.

\section{Results}

\subsection{Cutpoint selection}

In all maximum likelihood analyses, the bootstrap support requirement of $70$\% resulted in greater overlap with the reference set. Under the maximum patristic distance scheme of \cite{Brenner2007}, we found that a distance requirement of $7.7$\% maximised the correspondence (\gls{ari} = $0.91$). With ClusterPicker, requirements of either $6.8$\% or $7.7$\% were preferable (\gls{ari} = $0.91$). In PhyloPart, a median within-cluster patristic distance requirement of $0.03$ resulted in the largest overlap with the reference (\gls{ari} = $0.98$). Finally, in the Bayesian analysis, in addition to a posterior probability requirement of $1$, we determined that a $6.8$\% or $7.7$\% requirement for maximum within-cluster Hamming distances were equivalent (\gls{ari} = $0.91$). Except for \cite{Prosperi2011}, published clustering analyses tend to rely on more restrictive distance requirements and so, in cases where several distance requirements were equivalent, we picked the smallest one.

It is no surprise that the proposed schemes resulted in similar choices of cutpoints. Cluster estimation based on the consensus tree computed from the Bayesian tree search relies on ClusterPicker, just like one of the ML-bootstrap approaches. Normally, clusters found through a Bayesian analysis agree substantially with those obtained through a ML-bootstrap approach. Also, ClusterPicker uses maximum within-cluster pairwise distances, which provide a rough approximation of patristic distances. It follows that tuning ClusterPicker to produce estimates in line with those from the method of \cite{Brenner2007} should be feasible.

\subsection{Estimates comparison} \label{section:estimatesComparison}

\begin{table}
  \resizebox{\textwidth}{!}{ 
    \begin{tabular}{rrrrrrr}
      \hline
      & ML + CP & ML + PhyloPart & ML + Max. pat. dist. & MrBayes + CP & Gap Procedure & DM-PhyClus \\
      \hline
      ML + CP & 1.00 & 0.92 & 0.93 & 0.94 & 0.83 & 0.65  \\
      ML + PhyloPart & 0.92 & 1.00 & 0.91 & 0.86 & 0.88 & 0.68 \\
      ML + Max. pat. dist. & 0.93 & 0.91 & 1.00 & 0.88 & 0.83 & 0.66 \\
      MrBayes + CP & 0.94 & 0.86 & 0.88 & 1.00 & 0.84 & 0.64 \\
      Gap Procedure & 0.83 & 0.88 & 0.83 & 0.84 & 1.00 & 0.72 \\
      DM-PhyClus & 0.65 & 0.68 & 0.66 & 0.64 & 0.72 & 1.00 \\     
      \hline  
    \end{tabular}
  }
  \caption[Adjusted Rand index for the overlap between the cluster estimates obtained from the different methods.]{\textbf{Adjusted Rand index for the overlap between the cluster estimates obtained from the different methods.} CP stands for ClusterPicker.} \label{table:correspondenceARI}
\end{table}

We first compare optimal estimates from all methods with the \gls{ari}, cf. Table \ref{table:correspondenceARI}. We observed the largest overlap between the partitions resulting from the \textit{ML + maximum patristic distance} and \textit{ML+ClusterPicker} methods (\gls{ari} = $0.94$). On the other hand, we obtained the smallest overlap between estimates suggested by the MrBayes+CP and DM-PhyClus methods (\gls{ari} = $0.64$). DM-PhyClus produced the most distinctive set of clusters, with overlap with clusters from the other methods ranging from $0.64$ and $0.72$. The larger correspondence with the Gap Procedure estimate is not surprising, since both methods define clusters in terms of their separation from other clusters.

\begin{figure}[ht]
  \centering
  \includegraphics[width = 12cm]{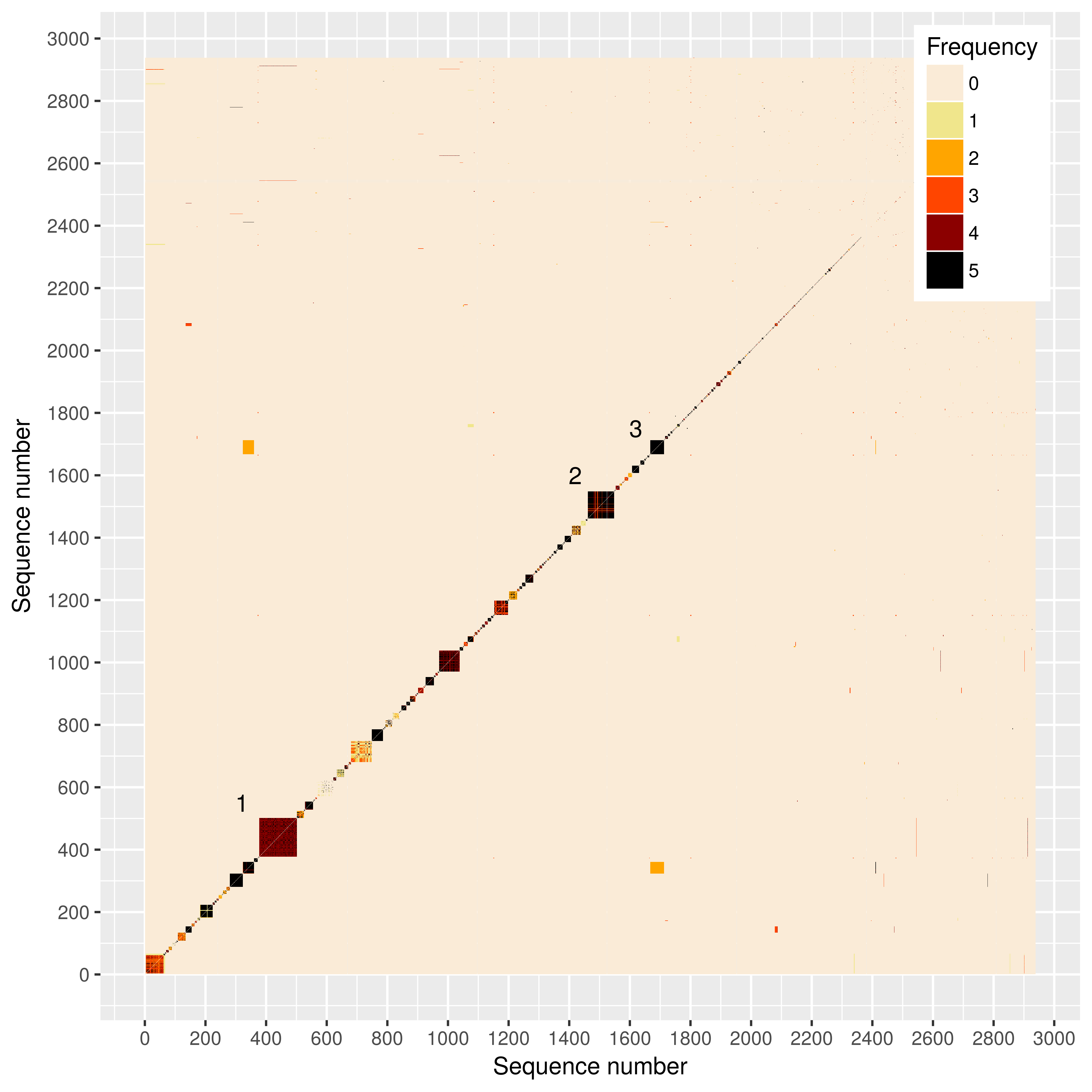}
  \caption[Heat map showing the frequency at which sequences co-clustered across methods.]{\textbf{Heat map showing the frequency at which sequences co-clustered across methods.} The $x$ and $y$ axis represent the $2,938$ sequences that were found to be non-singletons by at least one of the methods.}  
  \label{fig:overlapHeat}
\end{figure}

We represent graphically the correspondence between the different estimates in Figure \ref{fig:overlapHeat}. The heat map, showing the $2,938$ sequences found to co-cluster with at least one other sequence by at least one of the methods, reveals $11$ moderately-sized clusters. The largest rectangle, marked ``1'' in the figure, matches roughly one of the reference clusters, and is of size $\approx 125$. The earliest sequence in the cluster was collected on August 13, 2002 and was a \gls{phi}, and the latest sequence, also corresponding to a \gls{phi}, was obtained on December 23, 2015. The \gls{map} estimate of DM-PhyClus, on the other hand, split this cluster into $14$ components, including three clusters of size $37$, $36$, and $14$, respectively, and $7$ singletons. Instead of the \gls{map} estimate, we could have derived the so-called \textit{linkage estimate} from the chain results \cite{Villandre2017arxiv}. Broadly speaking, the linkage estimate proposes clusters by partitioning the sample into subsets of sequences that co-cluster often across iterations in the chain. We present a more detailed description in Supplementary Material S4. The linkage estimate ends up more in line with the other estimates: it contains $12$ components, with $3$ large clusters of sizes $37$, $36$, and $30$, and $7$ singletons.


The second largest cluster, represented by the mostly black block on the right, marked ``2'' in the figure, comprises $87$ sequences, and is also part of the reference set. All methods agree more or less that it indeed represents a transmission cluster. It comprises sequences sampled from May 11, 2004 (chronic untreated) to December 14, 2015 (\gls{phi}), which highlights its durability. The moderately-sized black block to its immediate right, marked ``3'', also stands out. Its $45$ sequences, also found in the reference set, co-cluster according to all the methods. Its first sequence was collected on January 11, 2012 and its last, on April 8, 2015. Two methods, \textit{MrBayes + ClusterPicker} and \textit{ML + ClusterPicker} added to that cluster an extra $38$ sequences, as evidenced by the light orange rectangle underneath it.


\begin{figure}[ht]
  \centering
  \includegraphics[width = 14cm]{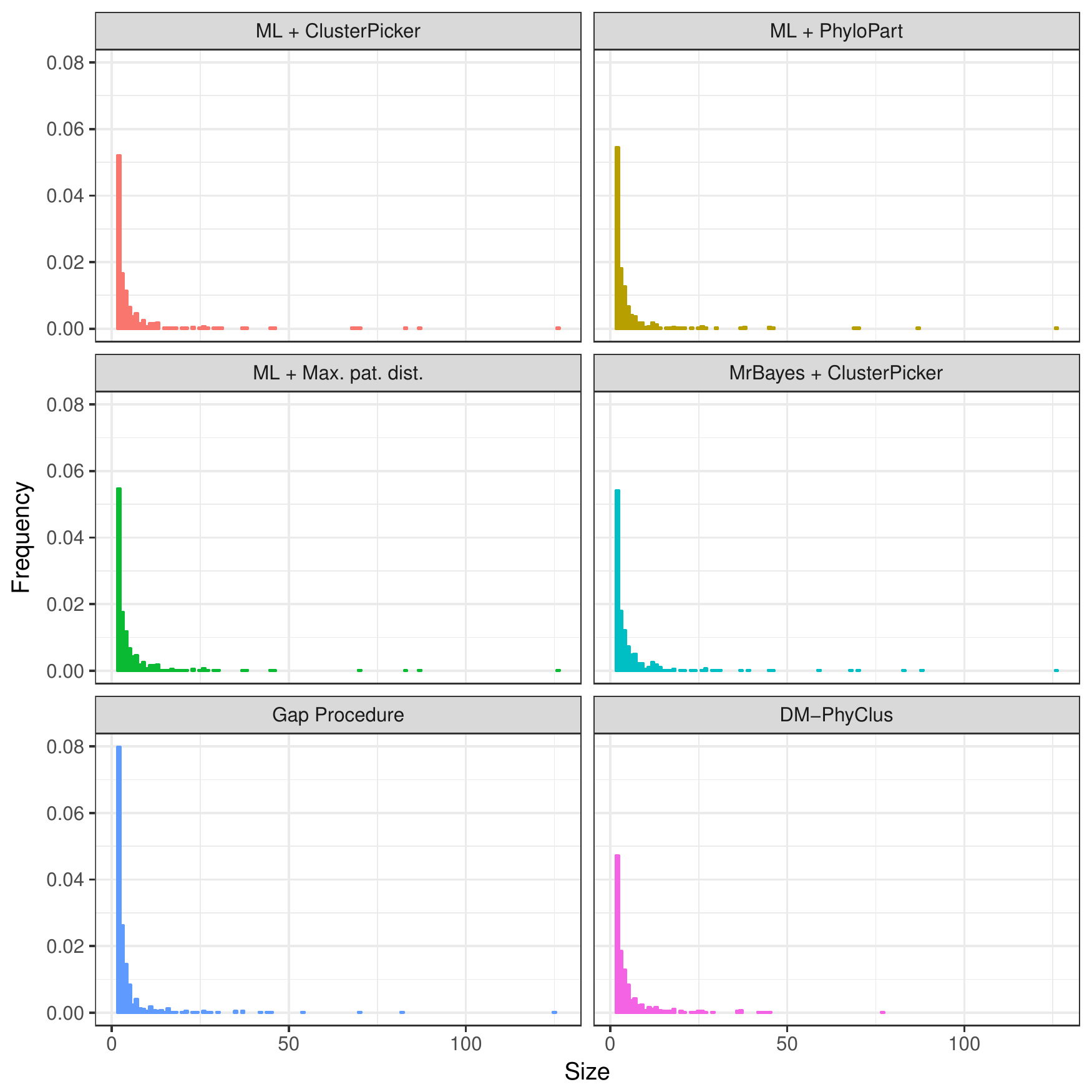}
  \caption[Truncated cluster size distributions for the preferred estimate across methods.]{\textbf{Truncated cluster size distributions for the preferred estimate across methods.} We refer to the figure in section \ref{section:estimatesComparison}, that focuses on summary statistics for the obtained cluster estimates, in order to highlight their differences. To improve readability, we removed the bars corresponding to singletons.}  
  \label{fig:clusSizeDists}
\end{figure}

\begin{table}
  \resizebox{\textwidth}{!}{ 
  \begin{tabular}{rrrrrrr}
    \hline
    & ML + ClusterPicker & ML + PhyloPart & ML + Max. pat. dist. & MrBayes + ClusterPicker & Gap Procedure & DM-PhyClus \\
    \hline
    Mean clus. size & 2.29 & 2.08 & 2.19 & 2.48 & 2.33 & 2.11 \\
    Mean (no singletons) & 6.01 & 5.53 & 5.62 & 5.96 & 4.80 & 5.62 \\
    Median (no singletons)  & 3.00 & 3.00 & 3.00 & 3.00 & 2.00 & 3.00 \\
    Max. clus. size & 126 & 126 & 126 & 126 & 125 & 77 \\    
    Num. singletons & 1205 & 1353 & 1261 & 1051 & 1035 & 1330 \\
    Num. clus. size $\geq$ 2 & 1621 & 1779 & 1696 & 1497 & 1592 & 1753 \\
    \hline
  \end{tabular}
  }
  \caption[Summary statistics for estimates returned by the different methods.]{\textbf{Summary statistics for estimates returned by the different methods.}} \label{table:clusSizesSummary}
\end{table}

Figure \ref{fig:clusSizeDists} presents truncated cluster size distributions derived from the preferred estimate from each method and Table \ref{table:clusSizesSummary} gives related summary statistics. Unsurprisingly, distributions obtained from the four conventional methods are very similar. Among those, the one for the conventional Bayesian estimate, labelled ``MrBayes + ClusterPicker'', stands out because of its thicker right tail. The distribution derived from the DM-PhyClus estimate is also distinctive, because of its much thinner right tail. Frequencies for singletons are not shown in the graphs for readability purposes. We found that ML + PhyloPart and DM-PhyClus had the highest proportions of singletons, each having approximately $36$\% of size $1$ clusters. On the other hand, the Gap Procedure and the conventional Bayesian estimate had the fewest singletons, with $28$\% of clusters having a single member. The Gap Procedure estimate, however, had much more transmission pairs than the other methods.

\subsection{Cluster growth assessment}

\begin{figure}[p]
  \centering
  \includegraphics[width = 12cm]{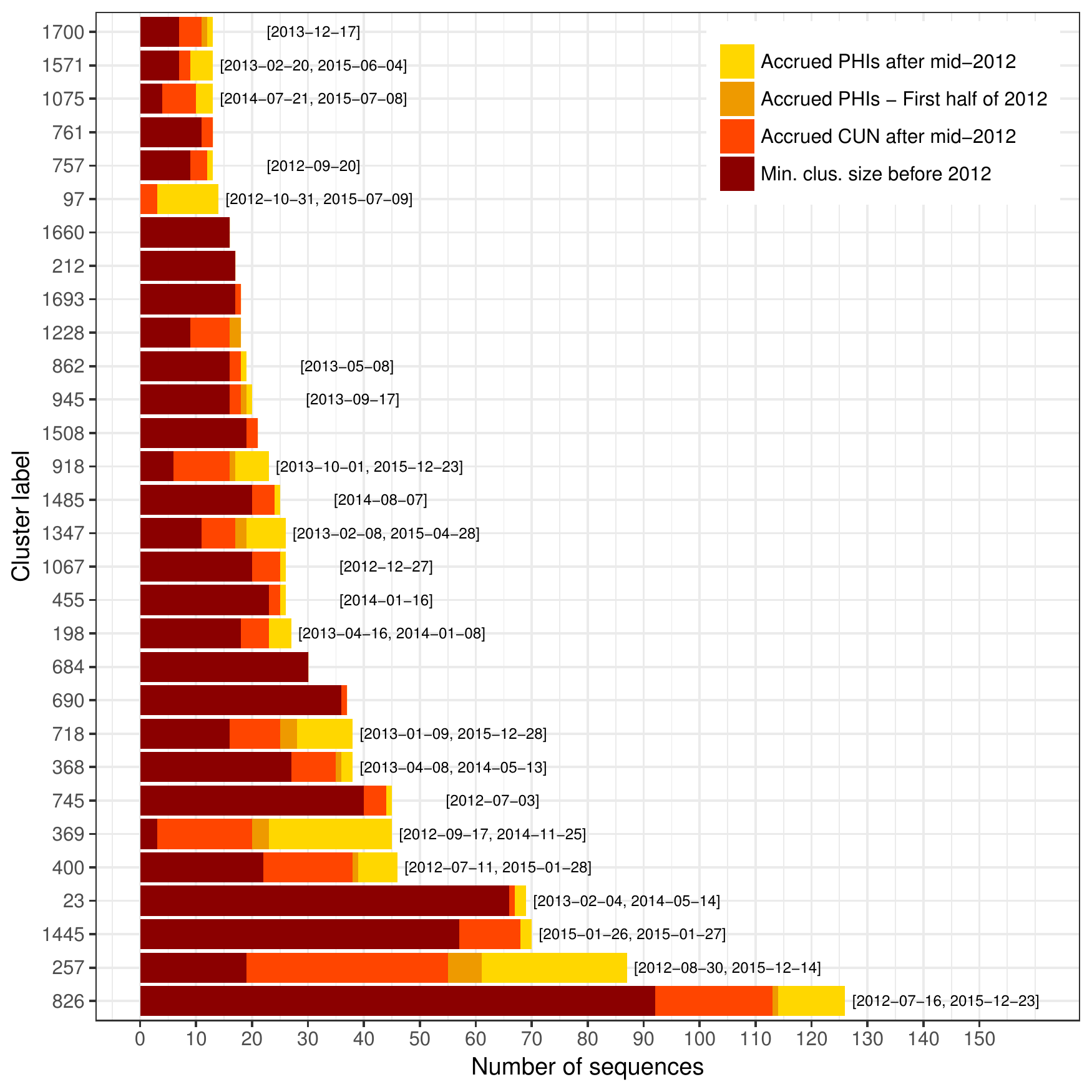}
  \caption[Bar plot showing the breakdown in membership for the 30 largest clusters in the ML + PhyloPart estimate.]{\textbf{Bar plot showing the breakdown in membership for the 30 largest clusters in the ML + PhyloPart estimate.} The labels at the end of each bar indicate the sequence collection dates for the first and last ``recent'' \glspl{phi} in the cluster, that is, recorded on or after July 1st, 2012. When there is only one such \gls{phi}, we display the corresponding collection date instead. We assume that all chronic cases recorded before July 1st, 2012 were infected prior to 2012. The dark red bar represents the ``minimum cluster size before 2012'' because several chronic cases diagnosed after July 1st 2012 were probably infected prior to 2012. Also, it is likely that several \glspl{phi} sampled in the first half of 2012 match with transmission events that occurred late in 2011.}  
  \label{fig:PHIbarPlotPhyloPart}
\end{figure}

Among the $957$ cases in the \gls{msm} risk group added to the database in the period ranging from January 1, 2012 to February 1, 2016, $304$ were \glspl{phi}. Of those \glspl{phi}, $254$ were sampled after June $30$, guaranteeing that the corresponding transmission events took place in 2012. According to the \textit{ML + PhyloPart} estimate, $50$ ($20$\%) of those $254$ \glspl{phi} are singletons, $23$ ($9$\%) are found in transmission pairs, and $153$ ($60$\%) belong to clusters of size five or more. In comparison, in the period ranging from July 1st 2008 to January 1st 2012, $319$ \gls{msm} cases diagnosed in the \gls{phi} stage were added to the database. After excluding all sequences sampled after January 1st 2012, we find that of those \glspl{phi}, $83$ ($26$\%) form singletons, $34$ ($11$\%) belong to transmission pairs, and $159$ ($50$\%) are part of clusters of size five or more. If we do not exclude the more recent sequences, we find that $79$ of the $319$ cases ($25$\%) are still singletons, which tends to indicate that the more recent \glspl{phi} tend to cluster more.

We represent the $30$ largest clusters, according to the \textit{ML + PhyloPart} estimate, in Figure \ref{fig:PHIbarPlotPhyloPart}. Those clusters include $126$ recent \glspl{phi}, split between $22$ clusters. Among the largest ten clusters, nine include at least one recent \gls{phi}. The largest cluster includes $12$ recent \glspl{phi}, while the second and third include $26$ and two, respectively. Cluster $369$ is noteworthy: despite its small size prior to 2012, it has grown quickly, with the addition of $22$ recent \glspl{phi}. Cluster $97$, on the other hand, is still small, but has not been recorded before. Each of those two clusters has a \gls{phi} recorded as late as the second half of 2015, indicating that they may still be expanding. Other conventional estimates and the Gap Procedure lead to similar conclusions, as can be seen in Supplementary Material S5. 

\begin{figure}[p]
  \centering
  \includegraphics[width = 12cm]{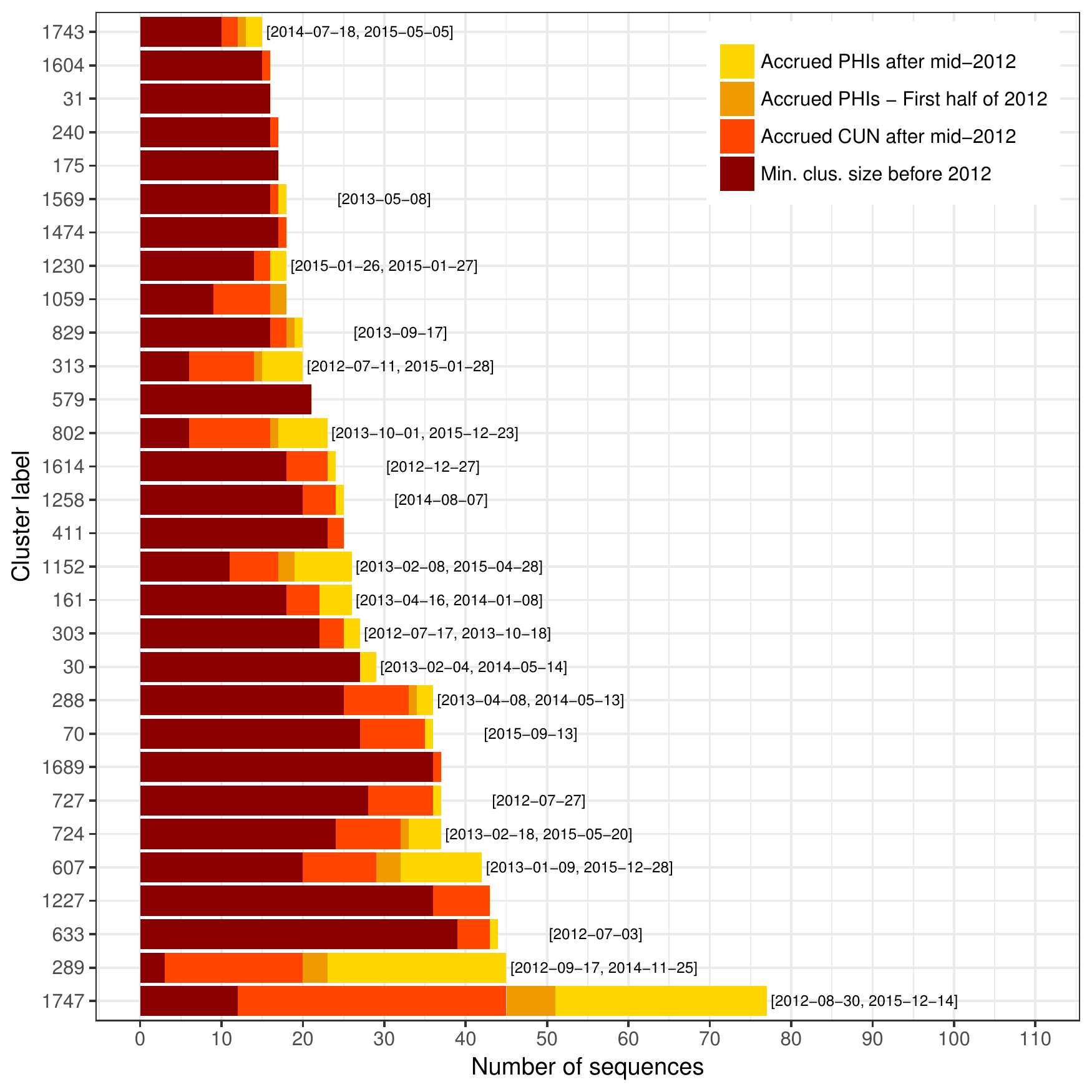}
  \caption[Bar plot showing the breakdown in membership for the 30 largest clusters in the DM-PhyClus estimate.]{\textbf{Bar plot showing the breakdown in membership for the 30 largest clusters in the DM-PhyClus estimate.} The labels at the end of each bar indicate the sequence collection dates for the first and last ``recent'' \glspl{phi} in the cluster, that is, recorded on or after July 1st, 2012. When there is only one such \gls{phi}, we display the corresponding collection date instead. We assume that all chronic cases recorded before July 1st, 2012 were infected prior to 2012. The dark red bar represents the ``minimum cluster size before 2012'' because several chronic cases diagnosed after July 1st 2012 were probably infected prior to 2012. Also, it is likely that several \glspl{phi} sampled in the first half of 2012 match with transmission events that occurred late in 2011.}  
  \label{fig:PHIbarPlotDM}
\end{figure}

The partition produced by DM-PhyClus is different, but leads to similar conclusions, as shown in Figure \ref{fig:PHIbarPlotDM}. Of the $30$ largest clusters, $20$ include at least one recent \gls{phi}. The largest cluster overlaps largely with cluster $257$ in Figure \ref{fig:PHIbarPlotPhyloPart} and includes $26$ recent cases, while the second and third largest include $22$ and $1$, respectively. The fifth largest cluster includes $10$ recent \glspl{phi}, out of $42$ members, also hinting at considerable growth. 


\section{Discussion}

\subsection{Summary}

In this paper, we investigated clustering in a sample of $3,704$ \gls{hiv}-1 cases belonging to the men who have sex with men risk category. We compared estimates from six methods, four conventional approaches relying on a variety of cutpoints applied to phylogenetic estimates, and two additional recent approaches seeking to avoid cutpoints entirely, the Gap Procedure and DM-PhyClus. We found that estimates obtained from conventional methods were overall fairly similar. The estimate from DM-PhyClus involved a noticeably different, albeit not unreasonable, cluster size distribution. Unlike other methods however, DM-PhyClus provides a straightforward measure of co-clustering frequencies and so, we found that requiring a certain degree of co-clustering, through the \textit{linkage-xx} estimate, could change estimates for certain clusters. All estimates however produced a similar assessment of cluster growth in the period ranging from January 1st, 2012 to February 1st, 2016: nine of the ten largest clusters had grown in the selected period, three of those having accrued at least ten new cases. Further, we observed several emerging clusters.

\subsection{Limitations}

The study has several limitations. Cutpoint selection remains inherently subjective. Indeed, choosing cutpoints as to maximise overlap with a reference set does not guarantee that other clusters will be estimated well. Moreover, identifying a suitable reference set can be difficult. In our study, researchers involved in the Quebec HIV genotyping program proposed the set based on a curated clustering analysis they conducted. A different reference set might have led to different cutpoints. Several of the approaches we used did manage to recover the reference set very closely though, which suggests that it is not unrealistic.

DM-PhyClus, being a Bayesian method, rests on a number of prior assumptions, which are all more or less informative, and it follows that prior calibration is key. \cite{Villandre2017arxiv} suggest that estimates are reasonably robust to some prior assumptions, but it remains possible that a combination of very poorly chosen priors may result in misleading cluster estimates.

Reliable infection date estimates for cases diagnosed while in the chronic stage are unavailable and so, we could only obtain a lower bound for cluster growth between January 1st, 2012 and February 1st, 2016. The average time between seroconversion and diagnostic is between $2$ and $3$ years \cite{vanSighem2015}, and it follows that several chronic cases diagnosed after June $30$th might have been infected during the selected period. Estimating infection time from the fraction of ambiguous nucleotides in each sequence would have been possible\cite{Kouyos2011}, but the high standard deviation for such predictions would have limited their usefulness.

Because of the non-random sampling of cases, we cannot readily deduce from our estimates the population-level cluster size distribution. In the absence of covariate information, we cannot model the sampling process. If, for example, the probability for a case to be sampled correlates positively with cluster size, we might end up underestimating the size of smaller clusters and the number of singletons, and consequently, overestimating the contribution of clustering to the epidemic. Nevertheless, the results we presented provide good evidence of cluster growth, and that phenomenon alone warrants attention.

\subsection{Selecting a best transmission cluster estimate}

Determining which partition among the six proposed provides the most accurate representation of transmission clusters in the sample is difficult. The choice depends ultimately on our confidence in the assumptions of each approach, and on substantive knowledge. The agreement between estimates from the conventional approaches, although explained in great part by shared assumptions, is still a good sign. The moderately different partitions proposed by the Gap Procedure and DM-PhyClus are not erroneous: they result from the way the two methods define clusters. The two approaches also have additional aims and benefits. \cite{Vrbik2015} designed the Gap Procedure with scalability in mind, and \cite{Villandre2017arxiv} formulated DM-PhyClus in such a way that it could offer a straightforward measure of uncertainty around the returned clusters.

\subsection{Conclusion}

The existence of large transmission clusters is not only a feature of transmission of \gls{hiv}-1 among \glspl{msm} in Montreal: it has been observed across Europe and other regions of North America as well \cite{LeighBrown2011, Lewis2008, Bezemer2010, Avila2014}. The increasing size of sequence databases represents a considerable computational challenge, especially in the Bayesian framework, and so, scalability should be an essential feature of future clustering algorithms. We contend that methods that avoid cutpoint selection altogether are convenient and promising, and would benefit from further improvements. In addition to lightening their computational burden, adapting them to use time-stamp and covariate data, for example, would be a welcome extension. Further, methods designed to provide a clear measure of uncertainty for estimated partitions, like DM-PhyClus, would warrant more attention. Indeed, the strength of co-clustering between sequences within an inferred cluster may vary sizeably, and the separation between neighbouring clusters may not be very clear-cut. Such variability may be hard to measure rigorously under conventional phylogenetic clustering approaches. 

Phylogenetic surveillance of \gls{hiv} transmission among \glspl{msm} provides helpful clues for explaining the persistence of the epidemic. The portrait of clustering presented in this study suggests an ongoing contribution of transmission cascades to incidence, a finding that should inform public health strategies to reduce transmission rates.


\section*{Ethics approval and consent to participate}

The study has received ethics approval from the McGill Faculty of Medicine Institutional Review Board. Ethics approval for the Quebec HIV genotyping program was obtained from individual study sites, the Laboratoire de sant\'{e} publique du Qu\'{e}bec, and the Quebec Ministry of Health committee on confidentiality and access of information.  

\section*{Competing interests}

The authors declare that they have no competing interests.
  
\section*{Funding}

This work was supported by a training award from the Fonds de recherche du Qu\'{e}bec-Sant\'{e} (FRQS), funding from the Centre de Recherches Math\'{e}matiques (CRM), a Natural Sciences and Engineering Research Council of Canada (NSERC) Discovery Grant, and a Canadian Institutes of Health Research (CIHR) grant (CIHR HHP-126781). Additional funding was obtained from the FRQS-Flandres collaboration (dossier 202685).

\section*{Author's contributions}

LV wrote the article, performed the analyses. LV, AL, and DAS jointly formulated the study plan. AL and DAS suggested and reviewed analyses. BB and MR provided the \gls{hiv}-1 sequences. RI helped with the preliminary analyses of the sequences and suggested a reference set for tuning the clustering procedure.

\section*{Data availability}

The Quebec HIV genotyping program sequences are protected by strict confidentiality agreements and so, cannot be made publicly available. A small subset of sequences can be provided for verification purposes upon request. 

\section*{Acknowledgements}


The Quebec HIV genotyping program is sponsored by the Minist\`{e}re de la Sant\'{e} et des Services sociaux (MSSS) du Qu\'{e}bec and by the Fonds de recherche du Qu\'{e}bec (FRQ-S) R\'{e}seau SIDA/MI.

\bibliographystyle{vancouver}
\bibliography{thesisBiblio}

\section*{Supplementary Material S1: Cutpoint selection with a partial gold standard}

The reference set includes only a small fraction of sequences in the dataset, and acts therefore as a \textit{partial gold standard}. We select cutpoints for each method as to maximise overlap with that reference set. The lack of a reference solution for other sequences in the sample makes comparison with this standard non-straightforward. Let us assume we have a sample of size $10$, and that sequences $1$-$3$ and $4$-$6$ form two confirmed clusters, labelled $1$ and $2$, respectively. A representation for cluster membership in the full gold standard would be [1, 1, 1, 2, 2, 2, Not 1 or 2, Not 1 or 2, Not 1 or 2, Not 1 or 2]. To best quantify overlap with the full gold standard, in all partitions we test, all sequences that do not co-cluster with any element in the reference set are given a membership index equal to (Number of clusters found among sequences in the reference set + 1). The full gold standard is reformulated in such a way that all sequences outside the reference set are given index (Number of true clusters in the reference set + 1). In the example, the gold standard would be reformulated [1, 1, 1, 2, 2, 2, 3, 3, 3, 3]. Let's say a clustering algorithm returns configuration [1, 1, 2, 3, 3, 3, 3, 4, 4, 5]. To obtain the correct \gls{ari}, we would need to transform it into [1, 1, 2, 3, 3, 3, 3, 4, 4, 4].

\section*{Supplementary Material S2: MrBayes script}

\begin{quote}
  begin mrbayes;
  set autoclose=yes nowarn=yes;
  execute brennerCompleteData.nex;
  lset nst=6 rates=invgamma;
  outgroup AB254141;
  set beaglescaling=dynamic beaglesse=yes;
  mcmc nruns=2 nchains=4 ngen=3000000 samplefreq=500 diagnfreq=10000 printfreq=500 append=yes;
  sump relburnin=yes burninfrac=0.25;
  end;
\end{quote}

\section*{Supplementary Material S3: Tuning parameters used in the DM-PhyClus and Gap Procedure analyses}

\sectionmark{Supplementary Material S3: Tuning parameters}

\subsection*{DM-PhyClus}

\begin{itemize}
\item Number of discrete states for the within-cluster and between-cluster transition probability matrices: $20$,
\item Number of samples used to obtain transition probability matrices: $100,000$,
\item Radius around mean within-cluster and between-cluster branch length estimates: $25$\%,
\item Discrete gamma distribution parameter: $1$,
\item Bootstrap and distance requirements for initial cluster estimate: $90$\%, $0.045$,
\item Limiting probabilities: $(A = 0.38, T = 0.24, C = 0.16, G = 0.21)$,
\item Rate matrix $Q$:
$$
\begin{bmatrix} 
    -0.8891 & 0.0659 & 0.1324 & 0.6908 \\
    0.1047 & -0.7205 & 0.5477 & 0.0681 \\
    0.3096 & 0.8069 & -1.1801 & 0.0636 \\
    1.2540 & 0.0779 & 0.0494 & -1.3812
\end{bmatrix}
$$ 
\item Shape parameter for concentration parameter prior: $500$,
\item Scale parameter for concentration parameter prior: $0.2$,
\item Poisson rate for weight applied to the cluster membership vector prior: $2368$,
\item Number of iterations: $220,000$.
\end{itemize}

\subsection*{Gap Procedure}

\begin{itemize}
\item Threshold for largest gap search: $90$\%.
\end{itemize}

\section*{Supplementary Material S4: The linkage estimate}

We obtain the linkage estimate by first projecting each cluster membership vector produced by DM-PhyClus as an unweighted undirected network graph, where each sequence is represented by a vertex, and an edge between any two vertex implying co-clustering between the corresponding sequences. For example, cluster membership vector [$1$, $1$, $1$, $2$, $2$, $2$] would translate as a graph with six vertices, split between two disjoint components, each of those being a \textit{fully-connected} graph. In other words, all vertices within each component are inter-connected. We can express an unweighted undirected network graph with an \textit{adjacency matrix}, a symmetric matrix with as many rows and columns as vertices, with a $1$ ($0$) at position $(i,j)$ indicating a connection (no connection) between vertices $i$ and $j$. Elements on the diagonal are set to $0$.

Once we have adjacency matrices for all cluster membership states visited by the chain, we average all the matrices element-wise, resulting in an adjacency matrix for a \textit{weighted} undirected network. Values in that matrix, all between $0$ and $1$, indicate the strength of the association between any two sequences. We then run the \textit{walktrap algorithm} on the corresponding graph to identify \textit{communities} \cite{Pons2005}. Communities are sets of vertices that are a lot more interconnected than would be expected from chance alone. The walktrap algorithm works by performing a large number of short random walks on the graph. It starts at a random vertex, and jumps to neighbouring vertices a fixed number of times. It is based on the principle that a short random walk starting in a community is more likely to end up in the same community, because of the high degree of interconnectedness between its vertices. The algorithm then outputs an estimate of community structure in the form of a vector of arbitrary community labels, which corresponds to the desired linkage estimate.

\pagebreak

\section*{Supplementary Material S5: Additional bar plots depicting cluster growth between January 1st, 2012 and February 1st, 2016}

\sectionmark{Supplementary Material S5: Additional bar plots}

The bar plots in this section can be read like Figures \ref{fig:PHIbarPlotPhyloPart} and \ref{fig:PHIbarPlotDM}. The labels at the end of each bar indicate the sequence collection dates for the first and last ``recent'' \glspl{phi} in the cluster, that is, recorded on or after July 1st, 2012. When there is only one such \gls{phi}, we display the corresponding collection date instead. We assume that all chronic cases recorded before July 1st, 2012 were infected prior to 2012. The dark red bar represents the ``minimum cluster size before 2012'' because several chronic cases diagnosed after July 1st 2012 were probably infected prior to 2012. Also, it is likely that several \glspl{phi} sampled in the first half of 2012 match with transmission events that occurred late in 2011.

\begin{figure}[ht]
  \centering
  \includegraphics[width = 11cm]{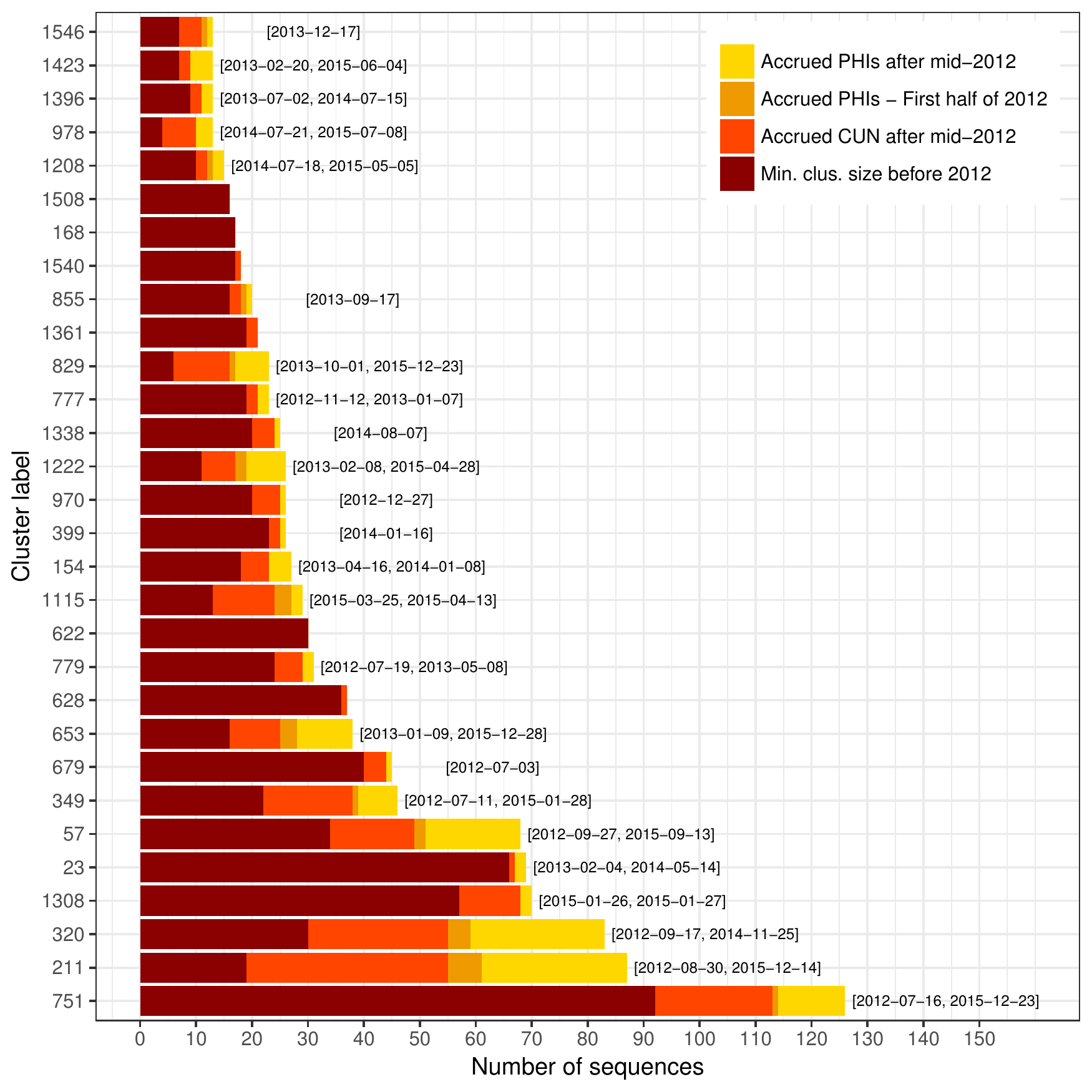}
  \caption[Bar plot showing the breakdown in membership for the 30 largest clusters in the ML + ClusterPicker estimate.]{\textbf{Bar plot showing the breakdown in membership for the 30 largest clusters in the ML + ClusterPicker estimate.} }  
  \label{fig:PHIbarPlotCP}
\end{figure}

\begin{figure}[ht]
  \centering
  \includegraphics[width = 11cm]{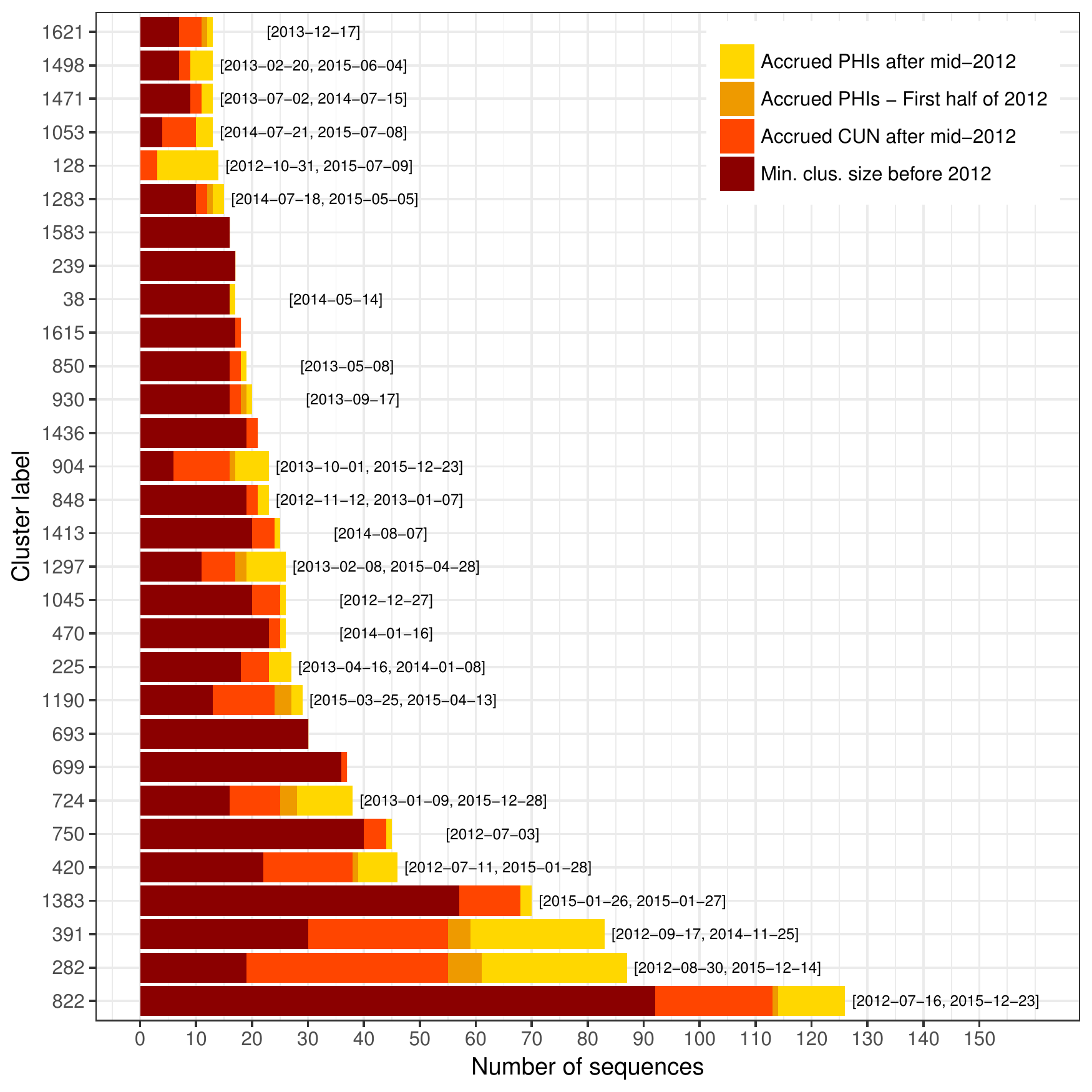}
  \caption[Bar plot showing the breakdown in membership for the 30 largest clusters in the ML + maximum patristic distance estimate.]{\textbf{Bar plot showing the breakdown in membership for the 30 largest clusters in the ML + maximum patristic distance estimate.}}  
  \label{fig:PHIbarPlotBrenner}
\end{figure}

\begin{figure}[ht]
  \centering
  \includegraphics[width = 11cm]{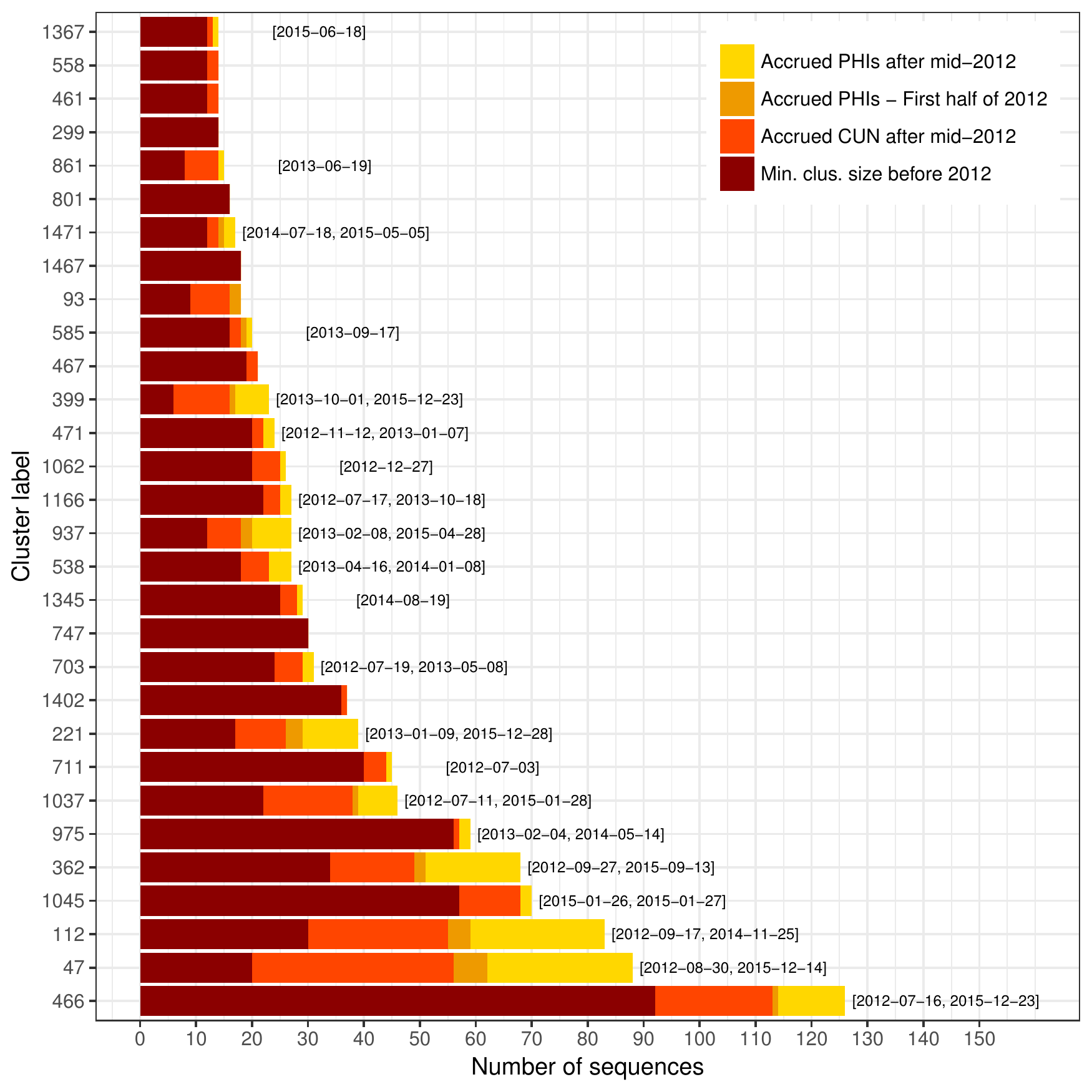}
  \caption[Bar plot showing the breakdown in membership for the 30 largest clusters in the MrBayes+CP estimate.]{\textbf{Bar plot showing the breakdown in membership for the 30 largest clusters in the MrBayes+CP estimate.}}  
  \label{fig:PHIbarPlotMrBayes}
\end{figure}

\begin{figure}[ht]
  \centering
  \includegraphics[width = 11cm]{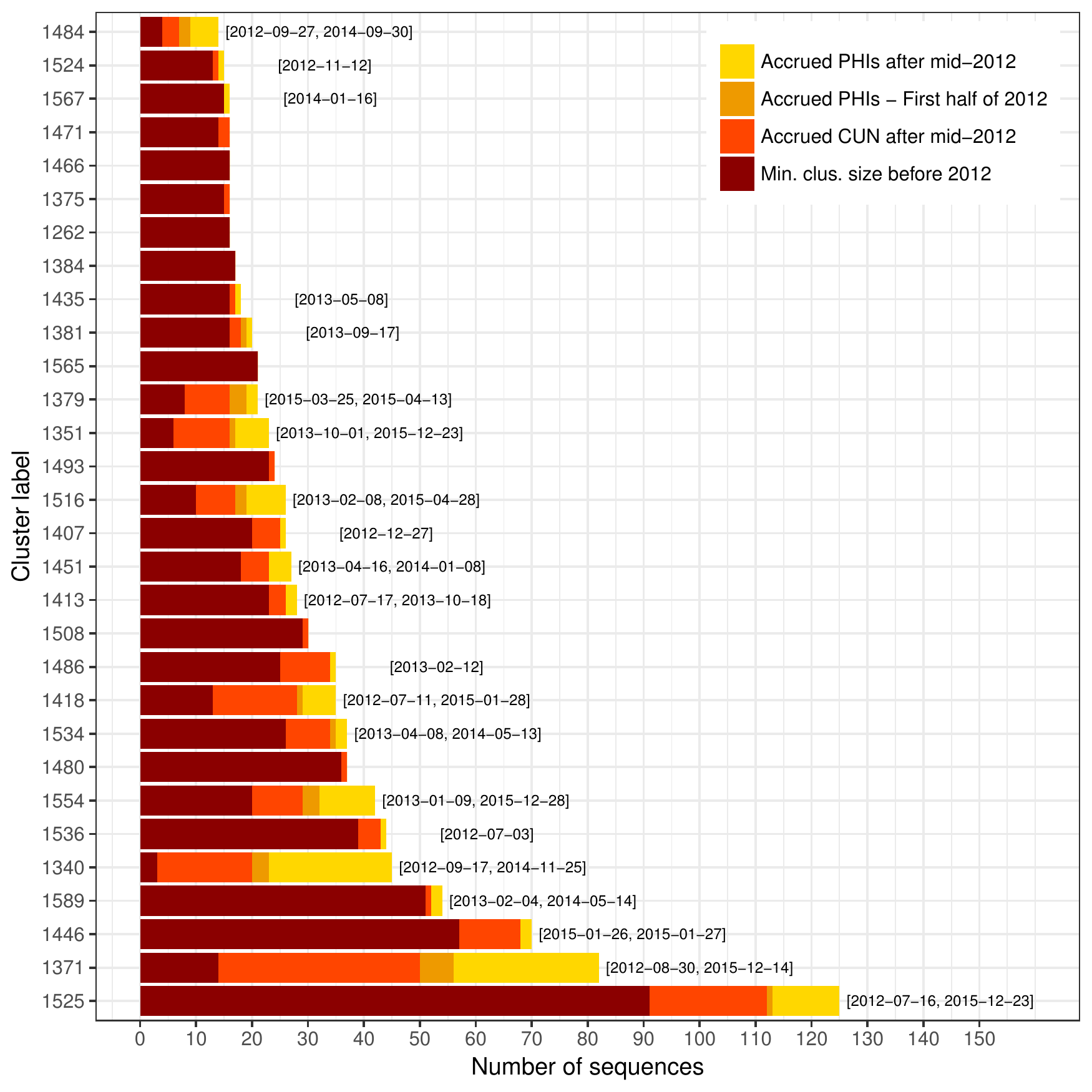}
  \caption[Bar plot showing the breakdown in membership for the 30 largest clusters in the Gap Procedure estimate.]{\textbf{Bar plot showing the breakdown in membership for the 30 largest clusters in the Gap Procedure estimate.}}  
  \label{fig:PHIbarPlotGap}
\end{figure}

\end{document}